\documentclass[letterpaper, 10 pt, conference]{ieeeconf}  
       
\IEEEoverridecommandlockouts                              
\title{\LARGE \bf Estimation of Heat Transfer Coefficient in Heat Exchangers from closed-loop data using Neural Networks}
\author{Ramachandran Anantharaman${}^a$, Carlos Gonzalez Rojas${}^a$, Luna Artemis van Leeuwen${}^{a}$ and Leyla Özkan${}^{a,b}$ 
\thanks{${}^a$ Eindhoven University of Technology, Eindhoven, The Netherlands. \vspace{.03in}\newline
${}^b$ Delft University of Technology, Delft, The Netherlands.\vspace{.03in} \newline Contacts: r.chittur.anantharaman@tue.nl, c.j.gonzalez.rojas@tue.nl, t.v.leeuwen@student.tue.nl, l.ozkan@tue.nl.
\vspace{.1in} \newline
This work was executed within the Institute for Sustainable Process Technology (ISPT) project: The Heat is On (THIO), and it is financed by  Topsector Energy of the Dutch Ministry of Economic Affairs and Climate Policy, executed by the Netherlands Enterprise Agency (RVO). The specific subsidy for this project is MOOI-subsidy 2020. }}

\usepackage{color}
\usepackage{float}
\usepackage{amsmath}
\usepackage{amsfonts}
\usepackage{amssymb}
\usepackage{esint}
\usepackage{hyperref}
\usepackage{varioref}
\usepackage{graphicx}
\newcommand{\al}{\alpha}

\newtheorem{theorem}{Theorem}
\newtheorem{proposition}{Proposition}
\usepackage{slashbox}
\usepackage{algorithm}
\usepackage{subcaption}
\usepackage{algorithmic}

\graphicspath{ {./figures/} }

\begin{document}

\maketitle
\thispagestyle{empty}
\pagestyle{empty}
 \begin{abstract}

  Heat exchangers (HEXs) play a central role in process industries for thermal energy transfer. Fouling, the gradual accumulation of solids on heat transfer surfaces, causes a time-varying decrease in the overall heat transfer coefficient $(U(t))$, significantly impacting the efficiency of heat transfer. Good estimation and modeling of fouling (the heat transfer coefficient) will lead to better fouling mitigation strategies. This study investigates the identifiability of the time-varying $U(t)$ in HEXs from closed-loop operational data, without external excitation of reference signals or knowledge of the controller parameters. We establish that while the complete system model cannot be identified under these given constraints, the time-varying heat transfer coefficient $U(t)$ remains identifiable. Further, we propose a neural network based architecture, called (Per-PINN), for estimation and modeling the heat transfer coefficient from the closed-loop system data. This Per-PINN model is shown to perform better than the existing Physics-Informed Neural Networks (PINN) based models for inverse parameter learning as it inherently fixes the underlying physical equations and learns only the time-varying parameter $U(t)$.
 \end{abstract}
\section{Introduction}
Heat Exchangers (HEXs) are widely used in various industries to transfer thermal energy between different fluids. Fouling in HEXs is the accumulation of particles on the heat transfer surface that inhibits the efficiency of heat transfer \cite{Bott_foulingbook,Rodriguez2007_Fouling}. This accumulation of particles increases the thermal resistance of the heat transfer surface and is reflected in a decrease of the heat transfer coefficient $U$. In industries, direct measurement of the change in $U$ is difficult and fouling is indirectly estimated by measuring process variables such as temperature, flow rate and pressure \cite{Diaz_2020}. Also, the fact that the heat transfer coefficient varies over time, poses a challenge in its estimation. 
Additional challenge arises when available data comes from a closed-loop system, where process variables are tightly regulated by controllers that compensate for fouling effects \cite{Patil2022_CL_Fouling}, thereby making it difficult to infer fouling from other process variables. 
Our objective in this paper is to estimate fouling in a heat exchanger (in particular, the heat transfer coefficient) where only the closed-loop data is available.


\subsection{Identification from closed-loop data}
\label{ss:CL-id}
\begin{figure}[h]
    \centering
    \includegraphics[scale = 0.25]{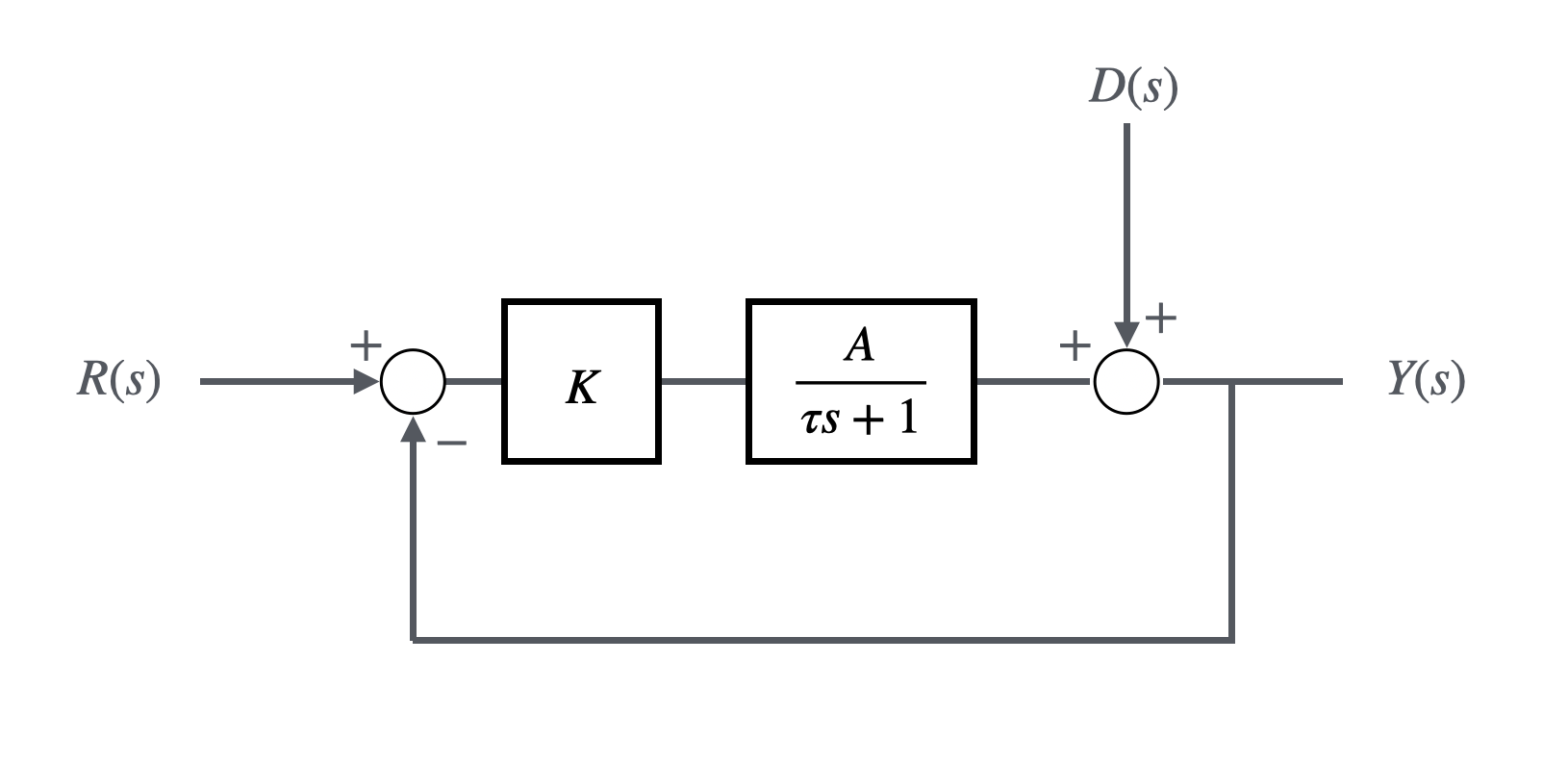}
    \caption{closed-loop Identification}
    \label{fig:CL_id}
\end{figure}

Parameter estimation and system identification have been a central focus in the field of process control, as a good model for the system helps in the design of better control strategies. In the context of HEXs, a good model for fouling or an estimate of $U$ can help with better monitoring of fouling, scheduling maintenance to remove fouling deposits leading to improved efficiency of the plant. An important difference between system and parameter identification is in the a priori knowledge about the system. In the former, we aim to find a dynamic (a transfer function or state space) model of a specific order which best fits the given data while in the latter, we aim to find certain unknown parameters which best fit the data under the constraints of the given model structure. Traditionally, identification is done with open-loop data, where the system is excited with a known signal, the corresponding output is measured, and these input-output data are used to develop models of the system. When data is available from a closed-loop operation, identification becomes non-trivial, and several methods of open-loop identification can not be extended to closed-loop data \cite{Ljung1999}. Furthermore, in the context of closed-loop identification, structural identifiability of open-loop system is not translated to the closed-loop system. For example, consider the closed-loop system as in Figure \ref{fig:CL_id}. 

 We are interested in finding the parameters $A,\tau, K$ from the closed-loop data. The closed-loop transfer functions are computed as 
\begin{align*}
    Y(s) = \begin{bmatrix} \frac{KA}{\tau s + KA +1} & \frac{\tau s +1}{\tau s + KA +1}\end{bmatrix} \begin{bmatrix} R(s) \\ D(s) \end{bmatrix}.
\end{align*}
 In the open-loop, the parameters $A$ and $\tau$ of the system can be uniquely identified, but in a closed-loop setting, a combined identification of process and controller in the above system is not possible, as $K$ and $A$ cannot be distinguished from the experiments. Although $A$ cannot be identified, it can be seen that $\tau$ can be uniquely identified through experiments. Through this simple example, we see that identifiability of open-loop systems does not carry forward to identifiability of closed-loop systems. In addition, in the case of a known controller gain $K$, the system parameters $A$ and $\tau$ are identifiable with excitations of $R(s)$ or $D(s)$. So, identifiability in a closed-loop setting is not only intrinsic to the system but also depends on the prior knowledge of the system and controller, as well as the access to excitations of different inputs. 
  \subsection{Identification of HEX from data}
In general, heat transfer inside the HEX between the two fluids can be modeled as a PDE, which, when discretized (into N spatial compartments), gives rise to a system of ODEs. In this work, we assume a lumped parameter model for the entire HEX.  
\begin{figure}[h]
    \centering
    \includegraphics[scale = 0.35]{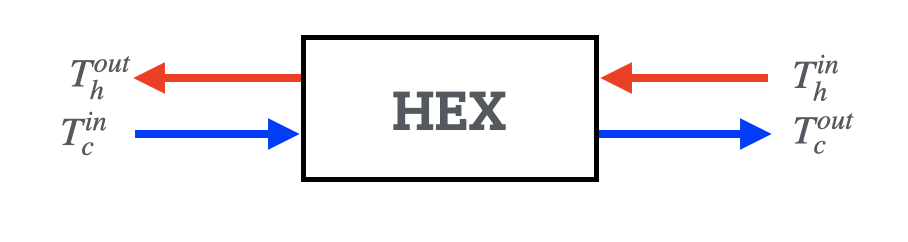}
    \caption{Heat Exchanger}
    \label{fig:HEX_OL}
\end{figure}

In a counterflow configuration of the HEX as shown in Figure \ref{fig:HEX_OL}, assuming that the fluid temperature is homogenized within the HEX and equal to the outlet temperature, the dynamics is given through the following ODE system \cite{Laszczyk_2017}

\scriptsize{\begin{align}
    \begin{aligned}
    \frac{dT_h^{out}}{dt} &= \frac{v_h}{V_h} (T_h^{in}(t) - T_h^{out}(t)) + \frac{U(t) A}{c_{ph}\rho_{h}V_h} (T_c^{out}(t) - T_h^{out}(t)) \\
    \frac{dT_c^{out}}{dt} &= \frac{v_c}{V_c} (T_c^{in}(t) - T_c^{out}(t)) + \frac{U(t) A}{c_{pc}\rho_{c}V_c} (T_h^{out}(t) - T_c^{out}(t)),
    \end{aligned}
    \label{eq:HEX_DE}
\end{align}}
\normalsize{}
where $T_h^{in}, T_h^{out}, T_c^{in}, T_c^{out}$ are the inlet and outlet temperatures $(K)$ of the hot and cold side respectively. $A$ ($m^2$) is the heat transfer area, $U(t)$ is the heat transfer coefficient ($W/ (m^2K)$), $v_i$ is the volumetric flow rates $(m^3/s)$, $V_i$ is volume $m^3$, $c_{pi}$ is the specific heat capacity ($J/(kg K)$) and $\rho_i$ is the density ($kg/m^3$) of hot $h$ and cold streams $c$, respectively.

\begin{figure}[h]
    \centering
    \includegraphics[scale = 0.35]{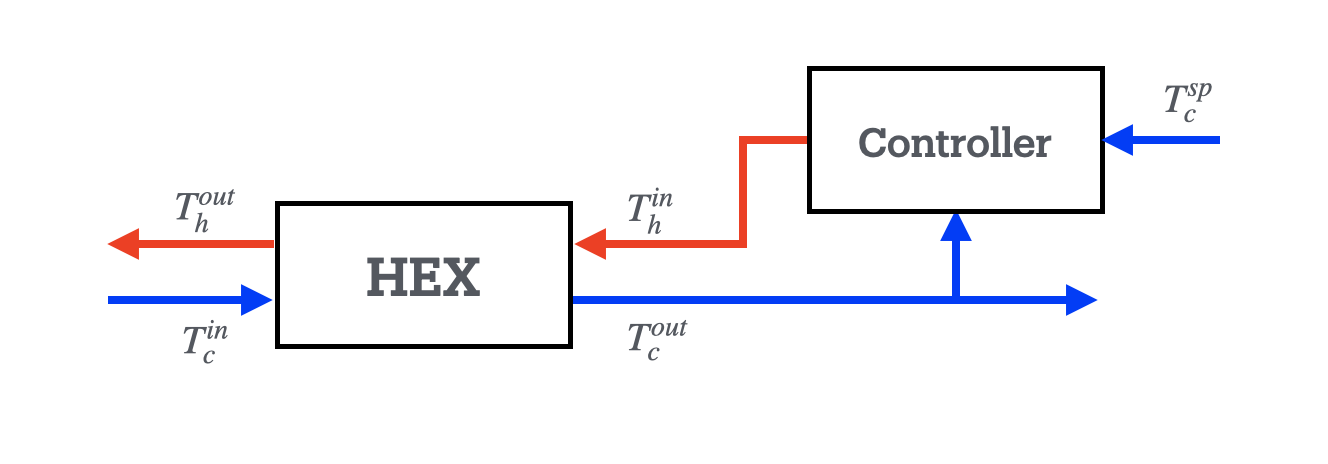}
    \caption{HEX in a closed-loop}
    \label{fig:HEX_CL}
\end{figure}

\begin{figure}[h]
    \centering
    \includegraphics[scale = 0.25]{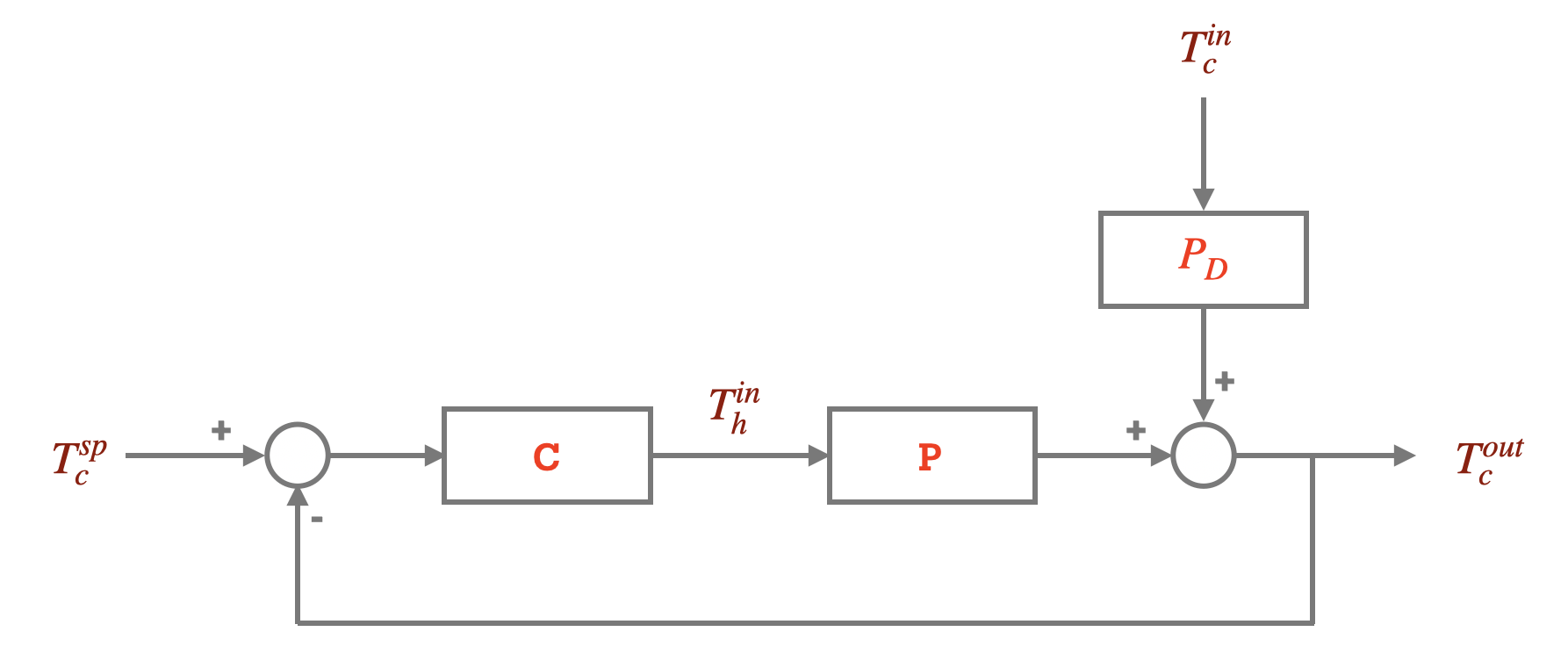}
    \caption{Block diagram of HEX in closed-loop}
    \label{fig:HEX_CL2}
\end{figure}

In an industrial setting, the HEX is generally operated in a closed-loop as shown in Figure \ref{fig:HEX_CL}, with temperature $T_c^{out}$ as the controlled variable.
We first study the identifiability of the HEX system. Converting the system to the standard closed-loop form as in Figure \ref{fig:HEX_CL2},  
with the inputs as $T_c^{sp}$ and $T_c^{in}$, the manipulated variable as $T_h^{in}$ and controlled variable as $T_c^{out}$, the transfer functions between these process variables are computed as 
\begin{align}
\begin{bmatrix} T_c^{out} \\ T^{in}_h 
\end{bmatrix} = \begin{bmatrix} \frac{CP}{1+CP} & \frac{P_D}{1+CP} \\ \frac{C}{1+CP} & -\frac{CP_D}{1+CP} 
\end{bmatrix} \begin{bmatrix} T^{sp}_c \\ T^{in}_c
\end{bmatrix}. 
\label{eq:TF_MV}
\end{align}
It can be seen that complete identification of the transfer functions $C$, $P$, and $P_D$ is possible if the two signals $T_c^{in}$ and $T_c^{sp}$ are allowed to be excited. If one of the signals is unavailable for excitation, the simultaneous identifiability of all the transfer functions is lost. For example, when you can only excite $T_c^{in}$, then only $P_D$ is identifiable while $P$ and $C$ are not. So, identification of the transfer functions (or equivalently, the complete differential equations governing the system) is not possible with closed-loop data when the inputs are not excited. This motivates the investigation on the identifiability of the parameter $U(t)$ of the ODE in \eqref{eq:HEX_DE}, rather than the complete ODE itself.  

\subsection{Problem formulation}
 The identification of $U(t)$ is the primary focus of this work, and we assume that $A$, $v_i$, $V_i$, $c_{pi}$ and $\rho_i$ are known a priori. In addition, the data used for identification come from a closed-loop operation, where neither the set point $T_c^{sp}$ nor the disturbance input $T_c^{in}$ is excited. We have access to all these process variables, along with $T_h^{out}$, the outlet temperature of the hot stream (see Figure \ref{fig:HEX_CL}). We further assume that the controller parameters are unknown but that the structure of the controller is known. The main objectives of this paper are as follows.
 
 In the context of limited or no access to excitation of the external inputs, we aim to 
 \begin{itemize}
     \item Given the ODE model obtained from the energy balance as in \eqref{eq:HEX_DE}, investigate the identifiability of the time-varying heat transfer coefficient $U(t)$ from the measurements. 
     \item Given the nonlinear dynamics of $U(t)$ depends on many internal and external factors, develop a neural network based framework to model the dynamics of the heat transfer coefficient (inverse parameter learning).
 \end{itemize}   

\subsection{Identification of the heat transfer coefficient using Neural Networks}
Apart from estimation of heat transfer coefficient $U$ with the data, it is also important that we model its dynamics so that it can be used for predicting fouling. Existing physical models for the dynamics of $U$ are highly nonlinear, indirect (as they model the change in thermal resistance) \cite{Fryer_foulingmodel1988, Ebert_foulingmodel1995}, and complete physical laws governing fouling are difficult to obtain. In addition, these models are sensitive to the chemical properties of hot and cold fluids, making generalization difficult for different fluids. In this context, we aim to leverage neural network based models for learning the dynamics of the heat transfer coefficient. The use of neural networks for the estimation of the heat transfer coefficient and fouling dynamics is also explored in these works \cite{Ilyunin2024_NNforFouling,Benyekhlef2021_NNforFouling, Sundar_2020}. All these works assume that the data used for the prediction of fouling is from an open-loop system without considering the action of the controller, and in contrast, we assume that the data is from a closed-loop system with no excitations on the external signals.

The fact that neural networks are universal functional approximators makes it an attractive choice for approximating unknown nonlinear functions and the use of such models for system identification leads to a class of models, commonly referred to as \textit{black box} models \cite{Sjoberg_1994}. In these models, a neural network is trained to minimize the data loss, which is formulated using the error between the true and predicted values (known as the prediction error) of the neural network. Recently, the use of Physics Informed Neural Networks (PINNs) \cite{raissi_2019} has attracted considerable attention in parameter estimation (also referred to as inverse learning) where unknown parameters are estimated to respect certain physical laws. These frameworks, predominantly classified under \textit{gray box} models, incorporates the physical laws corresponding to the mass and/or energy balance in the loss functions along with the prediction errors. These additional loss functions make the neural network more interpretable. One of the important shortcomings of PINN based formulations is their inability to handle hard constraints. Although there are ways to extend the PINN formulations for hard constraints \cite{lulu_2021}, these works add hard constraints on the boundary conditions rather than on the physical loss. 


\subsection{Organization of the paper}
The paper is organized as follows. The following section discusses (and establishes) the identifiability of the heat transfer coefficient $U(t)$ without the excitation of the reference signal $T_c^{ref}$ or the disturbance input $T_c^{in}$. For this analysis, we assume that the controller is a $P$ controller with an unknown gain $K_p$. Section \ref{sec:P-PINN} discusses the neural network based identification of $U(t)$. The proposed neural network structure for identification of $U(t)$ is termed Per-PINN, as the architecture is inspired by Physics Informed Neural Networks (PINN) based structures. Section \ref{sec:NumEx} provides various simulation studies in which the Per-PINN based identification is implemented and the heat transfer coefficient is learned and compares the results with traditional PINN based learning of $U(t)$.

\section{Identifiability of Heat Transfer Coefficient}
\label{sec:U-Id}
\normalsize{}

As we discussed in section \ref{ss:CL-id}, when the controller is unknown, the identifiability of a closed-loop system depends on the structure of the controller. In our discussions, we assume a proportional controller with (unknown) gain $K_p$ and the closed-loop system \eqref{eq:HEX_DE} can be written as

\scriptsize{\begin{align}
    \begin{aligned}
    \frac{dT_h^{out}}{dt} &= \alpha_h (K_p(T_c^{ref}(t)- T_c^{out}(t)) - T_h^{out}(t)) \\ 
    & \ \ \ + U(t) \beta_h (T_c^{out}(t) - T_h^{out}(t)) \\
    \frac{dT_c^{out}}{dt} &= \alpha_c (T_c^{in}(t) - T_c^{out}(t)) + U(t) \beta_c (T_h^{out}(t) - T_c^{out}(t)),
    \end{aligned}
    \label{eq:HEX_CL}
\end{align}}
\normalsize{}
where $\alpha_i = \frac{v_i}{V_i}$ and $\beta_i = \frac{A}{c_{pi}\rho_i V_i}$ for $i = h,c$, the hot and cold side respectively. As fouling increases, assuming that other parameters remain constant, the heat transfer coefficient $U(t)$ decreases. For a specific $T_c^{in}$, this decrease in $U(t)$ decreases the temperature $T_c^{out}$. Due to the presence of the controller that compares $T_c^{out}$ with $T_c^{sp}$, this decrease is compensated for by an increase in $T_h^{in}$. So, even in the situation where neither of the external signals is excited, there is a control action due to the change in $U(t)$, which is reflected in $T_h^{in}$. We measure the temperatures $T_c^{in}, T_c^{out}, T_h^{in}$ and $T_h^{out}$. In equation (\ref{eq:HEX_CL}), the unknown parameter $U(t)$ can be viewed as a time-varying parameter which drives the system.

Traditional methods for parameter identifiability assume that the parameter is constant (and hence an augmented state with no dynamics) and resort to observability rank conditions-based analysis for establishing identifiability \cite{Abarbanel_2009}. These results do not carry forward to parameters that are time-varying. In this context, we use the result of \cite{Martinelli_2022} to establish the identifiability of $U$. Assuming the constant parameter $K_p$ as an additional state of the system and $U(t)$ as the time-varying parameter, the augmented system dynamics corresponding to \eqref{eq:HEX_CL} is 
{\begin{align}
\label{eq:Id1}
\begin{aligned}
 \dot{x}(t) = g_0(x) + &f_1(x) T_c^{sp} + f_2(x) T_c^{in} + g_1(x) U(t)  \\
 y(t) &= \begin{bmatrix} 1 & 0 & 0 
 \\ 0 & 1 & 0 \end{bmatrix} x(t),
 \end{aligned}
\end{align}}
\normalsize{}
where 
\[
x = \begin{bmatrix} T_h^{out} \\ T_c^{out} \\ K_p \end{bmatrix}, \quad g_0 = \begin{bmatrix} -\al_h K_p T_c^{out} - \al_h T_h^{out} \\ -\al_c T_c^{out} \\ 0 \end{bmatrix},
\]
\[
f_1 = \begin{bmatrix} \al_h K_p \\ 0 \\ 0 \end{bmatrix} \quad f_2 = \begin{bmatrix} 0 \\ \al_c \\ 0 \end{bmatrix} \quad g_1 = \begin{bmatrix} \beta_h(T_c^{out} - T_h^{out}) \\ -\beta_c(T_c^{out} - T_h^{out}) \\ 0 \end{bmatrix}.
\]
Assuming the two outputs as $h_1 =T_h^{out}$ and $h_2 = T_c^{out}$, computing the Lie derivatives, we obtain
\begin{align*}
    \mathcal{O} &= \begin{bmatrix} h_1 \\ h_2 \\ L_{g_0} h_1 \\ L_{g_0} h_2 \end{bmatrix} = 
    \begin{bmatrix} T_h^{out} & 0 & 0 \\ 0 & T_c^{out} & 0 \\ -\al_h & -\al_h K_p & -\al_h T_c^{out} \\ 0 & -\al_c & 0 \end{bmatrix}.
\end{align*}
It can be seen that the rank of $\mathcal{O}$ is $3$ and hence the system is (locally) observable. Since the system is observable, it can be seen from \cite{Martinelli_2022, Martinelli_2023} that 
\begin{proposition}
The time-varying parameter $U (t)$ of \eqref{eq:Id1} is identifiable. 
\end{proposition}
\vspace{.1in}The proof of the result is a direct application of \cite[Theorem 1 and Remark 4]{Martinelli_2023} and is given in the Appendix for completeness. It is pertinent to note that the identifiability of $U(t)$ may depend on the choice of controller structure (PI, PID etc.), but we believe that these investigations are beyond the scope of this paper as the aim was to establish identifiability of time-varying $U(t)$ at least in simple controller configurations. 


\section{Per-PINN architecture and its application to parameter identification}
\label{sec:P-PINN}

In recent years, several physically regularized neural network architectures have been proposed for parameter and model identification. Among them, physics-informed neural networks (PINNs) \cite{raissi_2019} and neural ordinary differential equations (neural ODEs) \cite{chen2018neural} are the two dominant families of methods found in the literature. The fundamental difference between these methods is that PINNs are differential-based estimators, while neural ODEs are integration-based predictors. In this paper, we propose the use of the ODE structure known from the energy balance to reduce the complexity of the learning process. Since the goal is to identify the heat transfer coefficient $U(t)$, we implement a “Per-PINN” considering that we know all components of the ODEs except $U(t)$. This is advantageous compared to the class of PINN methods because we skip training a neural network to approximate the evolution of the model states and focus directly on learning a model for $U(t)$. Moreover, we exploit and preserve the structure of the known components in the ODE, which is different from the common black-box neural ODE implementations.

To understand the differences between PINN and Per-PINN we first define a general inverse problem and then connect it with the PINN and Per-PINN formulations. Consider the HEX system represented by the ODEs in \eqref{eq:HEX_DE}, which can be rewritten in a compact form as 

\begin{equation}
    \frac{dT^{out}_i}{dt} - F(T^{in}_i, T^{out}_i, U) = 0.
    \label{eq:system1}
\end{equation}

\begin{figure}[h]
    \centering
    \includegraphics[scale = 0.23]{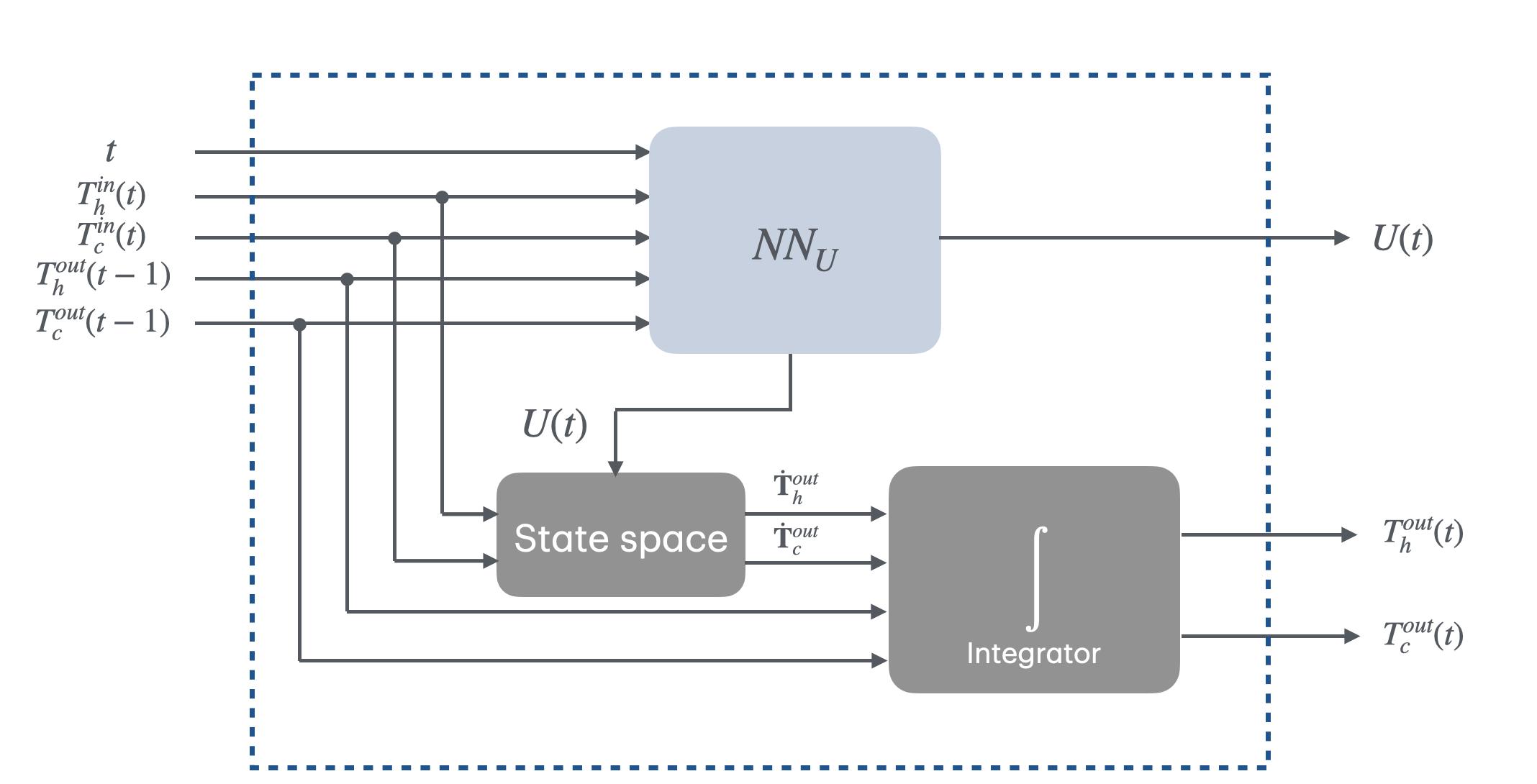}
    \caption{Per-PINN}
    \label{fig:Per-PINN}
\end{figure}

\begin{figure}[h]
    \centering
    \includegraphics[scale = 0.27]{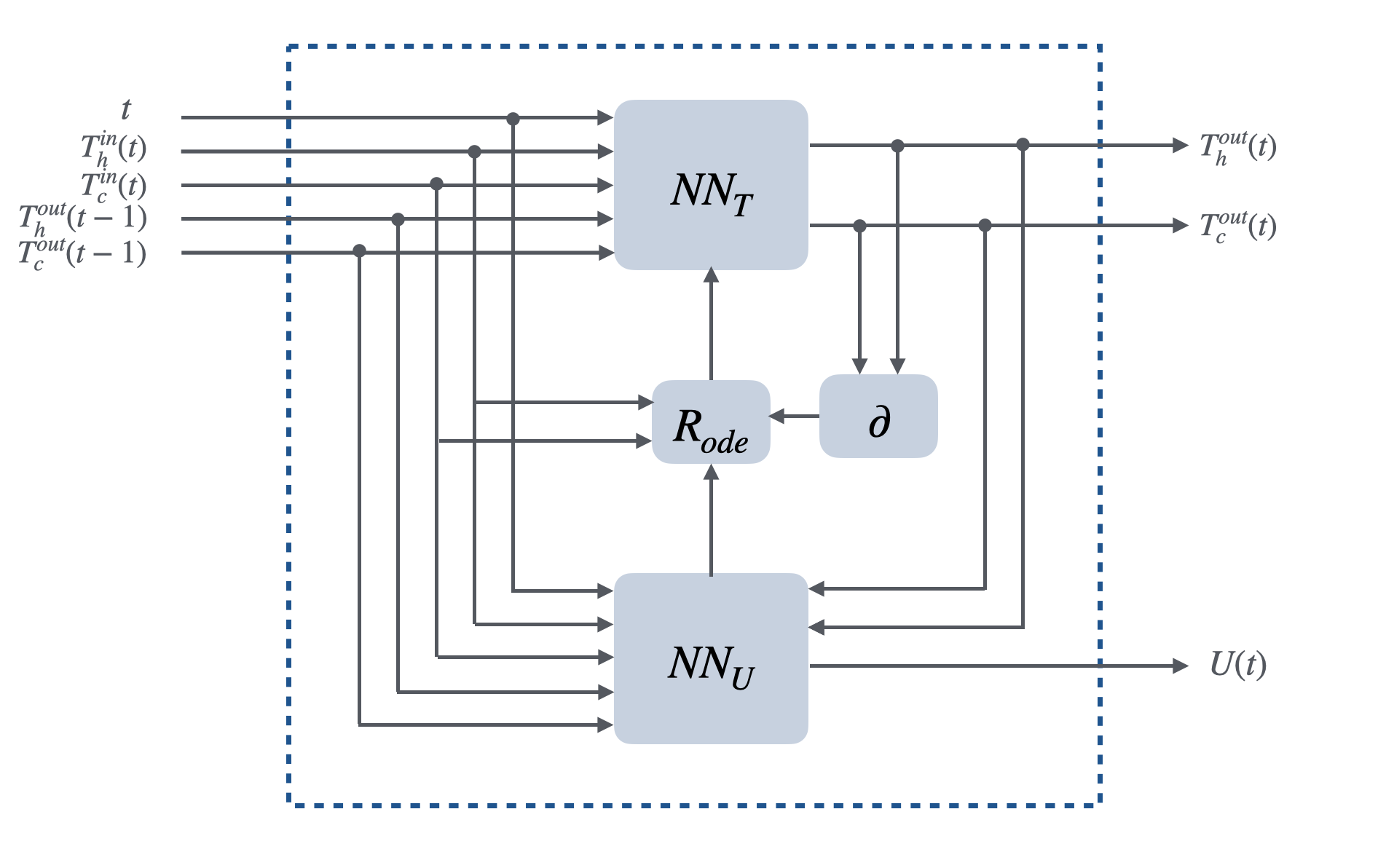}
    \caption{PINN}
    \label{fig:PINN}
\end{figure}

We define the residual of the ODE system \eqref{eq:system1} as

\begin{equation*}
    R= \frac{dT^{out}_i}{dt} - F(T^{in}_i, T^{out}_i, U),
\end{equation*}

and the inverse problem is defined as the following optimization problem

\begin{align*}
    \min_{\theta} \sum_{k=1}^{N} &\ell(T^{out}_i(k), \hat{T}^{out}_{i,\theta} (k)) \\
    \text{s.t.} &\quad R_{\theta} = 0.
\end{align*}

where $\theta$ represents the parameters of a function approximator and $\ell$ is a general loss function (e.g MSE, MAE). The objective is to find the optimal parameters $\theta$ that minimize the prediction error subject to the ODEs of the system. In standard PINNs, the function approximator is a neural network
\begin{align}
\hat{T}^{out}_{j,\theta} = NN_{\theta}(\cdot)
\end{align}
automatic differentiation is used to compute the residual $R_{\theta}$ and the equality constraint is treated as a weighted soft constraint
\begin{align}
    \min_{\theta} \sum_{k=1}^{N} \ell(T^{out}_j(k), \hat{T}^{out}_{j,\theta} (k)) + W_R  \ell (R_{\theta}).
\end{align}
In \cite{lulu_2021}, the authors analyzed the issues of using the residual as a soft constraint and proposed alternative solution strategies such as the penalty or the augmented Lagrangian methods. However, depending on the implementation, these methods can become computationally demanding. In contrast, the Per-PINN structure exploits the known ODE structure as follows
\begin{align*}
    \min_{U_{\theta}} &\sum_{k=1}^{N} \ell(T^{out}_j(k), \hat{T}^{out}_{j,U_{\theta}} (k)) \\
    \text{s.t.} \quad \hat{T}^{out}_j(U_{\theta}) &= ODESolver(F(T^{in}_j, T^{out}_j, U_{\theta})).
\end{align*}

where $U_{\theta} = NN_{\theta} (:)$ is learned using a neural network. Therefore, if the implementation allows backpropagation through $ODESolver()$, then the ODE system can be imposed as a hard constraint. A similar approach has been used in \cite{Gerben_2023, Liu_2024} and can be efficiently incorporated by using a neural ODE with the known structure. The Per-PINN and PINN architectures defined to learn the dynamic parameter $U(t)$ are described in Figures \ref{fig:Per-PINN} and \ref{fig:PINN} respectively.
 
\section{Numerical Results - Per-PINN for HEX}   
\label{sec:NumEx}
In this section, we discuss the simulation results of using the Per-PINN structure for recovery of the heat transfer coefficient $U(t)$ and compare it with a PINN-based structure. The data is generated from the model \eqref{eq:HEX_CL} with the following parameters $\alpha_h = 0.0996$, $\alpha_c = 0.0664 $, $\beta_h = 2.41\times 10^{-5}, \beta_c = 3.11 \times 10^{-5}$. In the system, the reference signal $T_c^{sp}$ is held constant at $391K$ and $T_c^{in}$ at $381K$. Furthermore, $T_c^{in}$ is assumed to be corrupted by a white Gaussian noise $\mathcal{N}(0,0.25)$. The temperatures $T_c^{in},T_c^{out}, T_h^{in}$ and $T_h^{out}$ are sampled every $50s$. The fouling rate was assumed to be affected by time and the average temperature within the HEX and a total of $30$ different runs were used to generate data of $3000s$ each, amounting to a total of $1800$ data points. In the plots, each time step corresponds to a sample of temperature data and we have $60$ data samples for each run. At the start of each run, the initial heat transfer coefficient $U$ was randomly assigned between $9500$ and $10500 \ W/(m^2K)$ (this mimics cleaning of the HEX before every run) and plotted the temperatures and estimated heat transfer coefficients over multiple ($5$) runs successively. 

\begin{figure}
\centering
\begin{subfigure}[b]{0.45\textwidth}
   \includegraphics[width=1\linewidth]{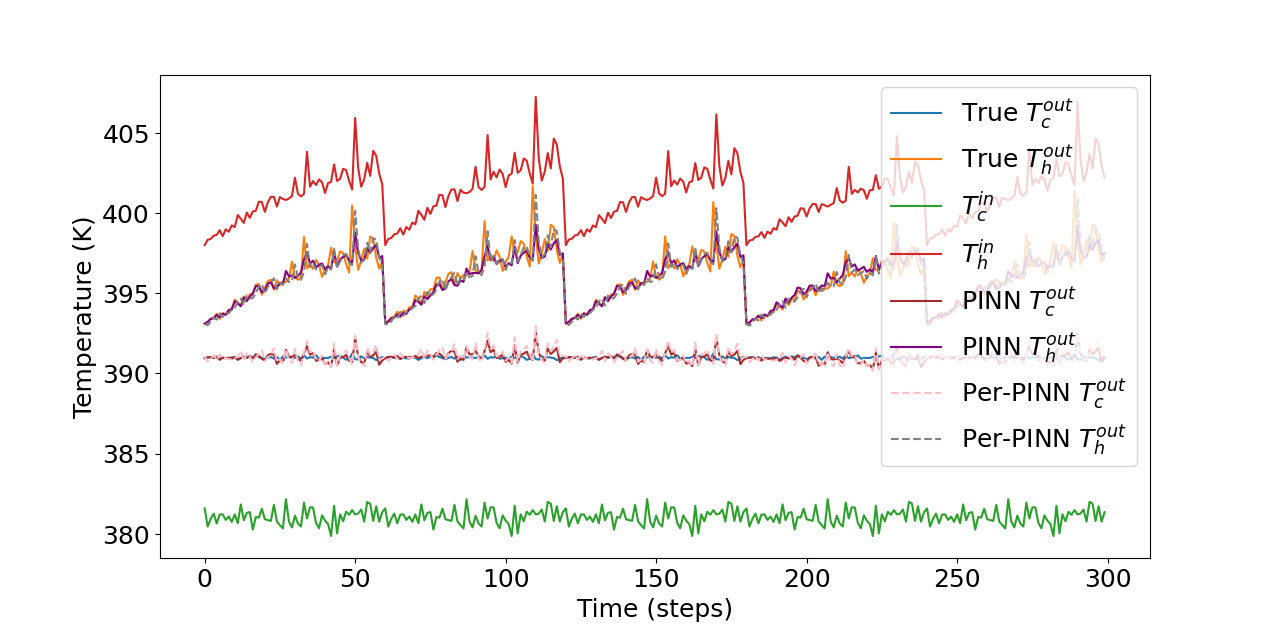}
   \caption{}
   \label{fig:v11} 
\end{subfigure}

\begin{subfigure}[b]{0.45\textwidth}
   \includegraphics[width=1\linewidth]{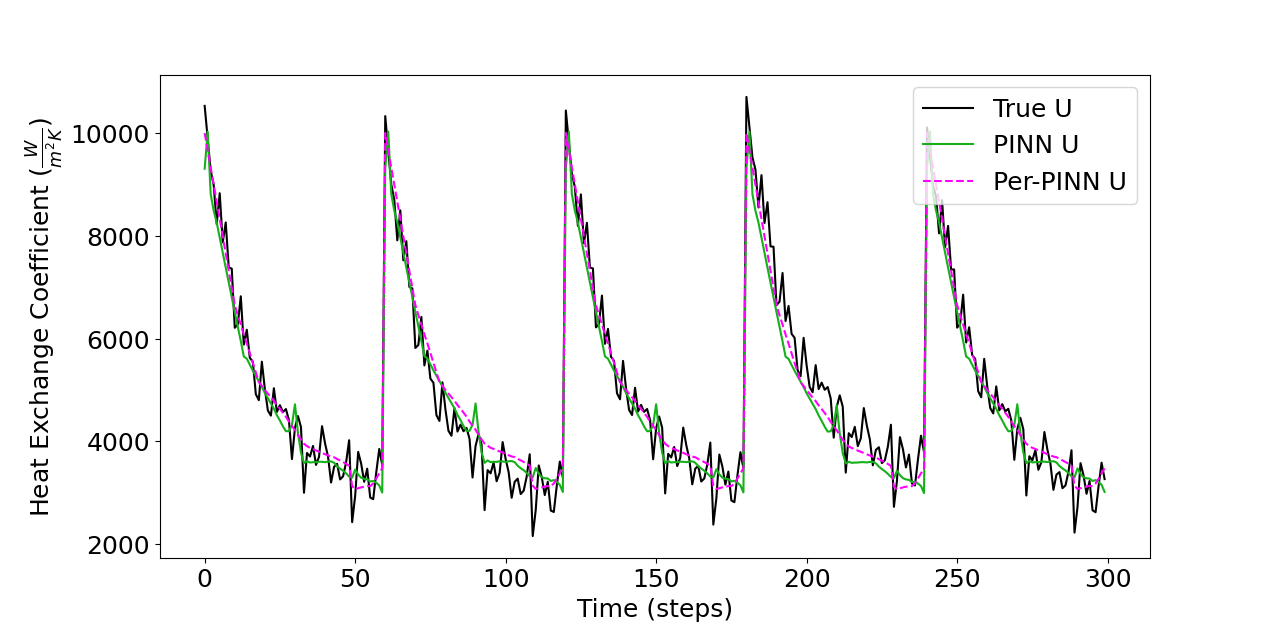}
   \caption{}
   \label{fig:v12}
\end{subfigure}

\caption[Validation of Per-PINN and PINN with closed-loop data]{(a) Prediction of $T_c^{out}$ and $T_h^{out}$ (b). Prediction of $U(t)$}
\label{fig:Validation1}
\end{figure}

\begin{figure}
\centering
\includegraphics[width=1\linewidth]{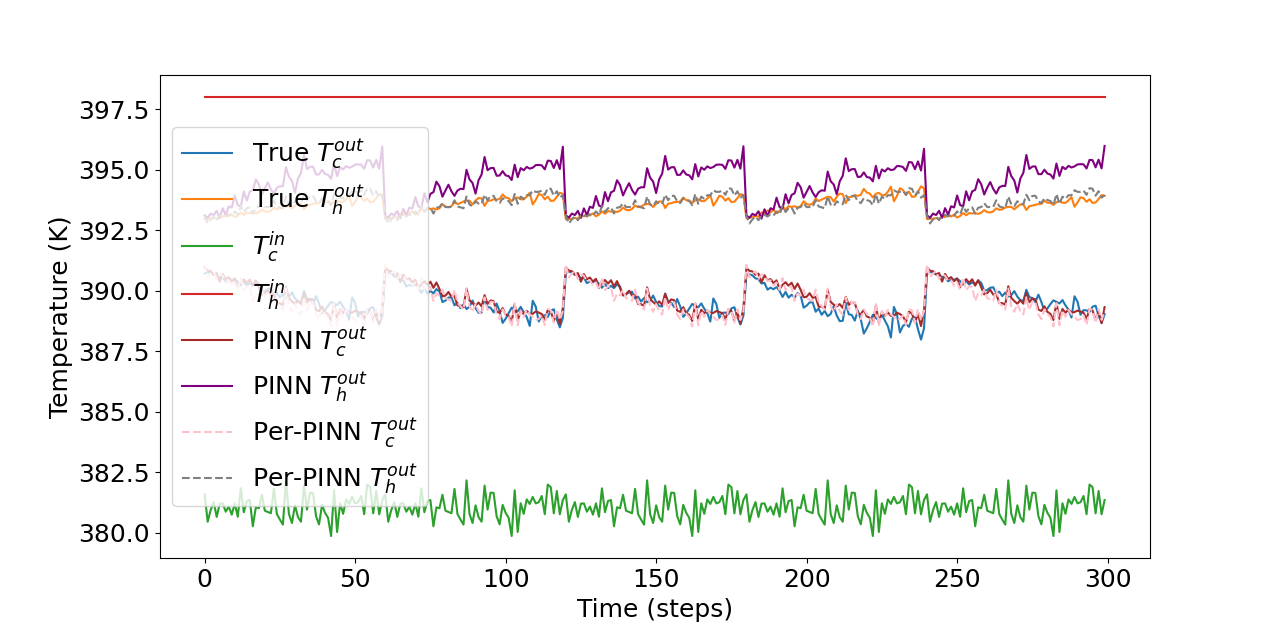}
   \caption{Testing of Per-PINN and PINN in prediction of $T_c^{out}$ and $T_h^{out}$ in open-loop}
   \label{fig:Validation2} 
\end{figure}

\begin{figure}
\centering
   \includegraphics[width=1\linewidth]{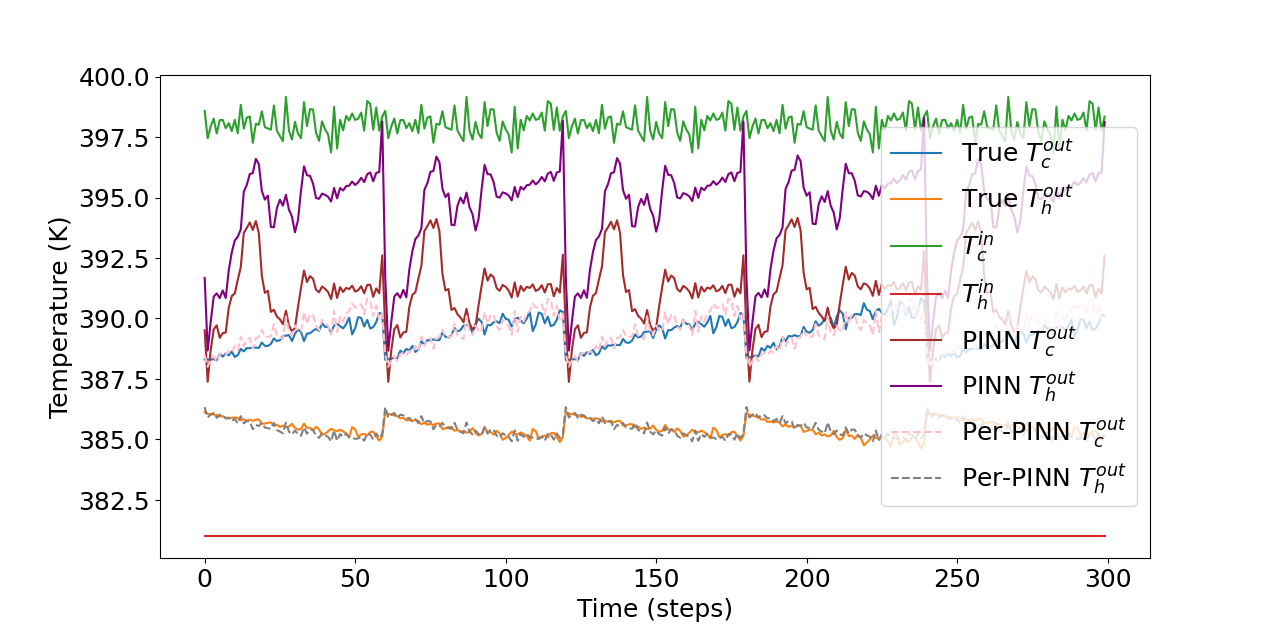}
   \caption{Testing of Per-PINN and PINN in prediction of $T_c^{out}$ and $T_h^{out}$ with switching of hot and cold streams}
   \label{fig:Validation3} 
\end{figure}

Regarding the training of neural networks: we choose a network of Per-PINN with $2$ layers and $75$ nodes, and for PINN we choose a network with $6$ layers and $40$ nodes. The loss function of Per-PINN was the data loss in the prediction of $T_c^{out}$ and $T_h^{out}$, while the loss function of PINN was the sum of data loss and the sum of the residuals of the ODE (scaled and) weighted equally. We used the \emph{deepxde} library \cite{Lu_deepxde2021} and performed a hyperparameter search to obtain the best performing PINN. We perform the following experiments:
\begin{enumerate}
    \item Using closed-loop data, we train Per-PINN to learn $U(t)$ and compare the results with a PINN. Once the networks are trained, we validate the estimates of $T_c^{out}$, $T_h^{out}$ for the test data, as well as the estimates of $U(t)$. It can be seen that both PINN and Per-PINN were able to predict the temperatures as well as $U(t)$ effectively. The results are shown in Figure \ref{fig:Validation1}. 
    \item We train Per-PINN and PINN with closed-loop data and learn $U(t)$. We then test the performance of both models using a new set of open-loop data generated from the same system parameters and initial conditions. In this context, we see that PINN performs poorly compared to Per-PINN in the estimation of $T_c^{out}$ and $T_h^{out}$, though their performances in estimating $U(t)$ are comparable. This supports the fact that, when the dynamics of the system is completely known, it is advantageous to use Per-PINN over PINN for learning the parameter alone. These results are shown in Figure \ref{fig:Validation2}.
    \item The final experiment is to compare Per-PINN and PINN models on how they respect the physical conservation laws properly when given completely unseen data. For this, we use the Per-PINN and PINN models trained using closed-loop data. A new set of test data is generated where the temperatures of the hot and cold streams are switched, and we verify whether the direction of heat transfer is reversed. We see that the Per-PINN model predicts the output temperatures properly, while the prediction of PINN is not accurate. This is illustrated in Figure \ref{fig:Validation3}. From this, we can conclude that when faced with non-ideal process data, which can be due to failures/errors in a different part of the plant or slightly different initial conditions, PINN can give you physically inconsistent predictions, while Per-PINN will always be physically consistent.  In fact, in a later example (Figure \ref{fig:Validation5}), PINN predicts $T_h^{out}$ more than $T_c^{in}$, the higher temperature in the reverse scenario, which is a violation of energy conservation. 
\end{enumerate}


The validation and testing errors are given in Table \ref{tab1} for both PINN and Per-PINN.  

\begin{table}
       \centering
     \scalebox{0.8}{
     \begin{tabular}{||c|c|c|c||}
    \hline
         &  Per-PINN & PINN (seed $1234$) & PINN (seed $4567$) \\ \hline 
         Closed-loop training  & $0.4985$ & $0.3489$ & $0.4186$\\ \hline
         Closed-loop validation & $0.4982$ & $0.3234$ & $0.3918$\\ \hline
         Open-loop testing & $0.1065$ & $0.5969$ & $0.5559$\\ \hline
         Stream switch testing & $0.1181$ & $45.68$ & $38.85$\\ \hline
    \end{tabular}}
    \caption{Comparison of validation and testing errors}
    \label{tab1}
   \end{table} 



\subsection{Discussions on the numerical study}
\begin{figure}
\centering
\begin{subfigure}[b]{0.45\textwidth}
   \includegraphics[width=1\linewidth]{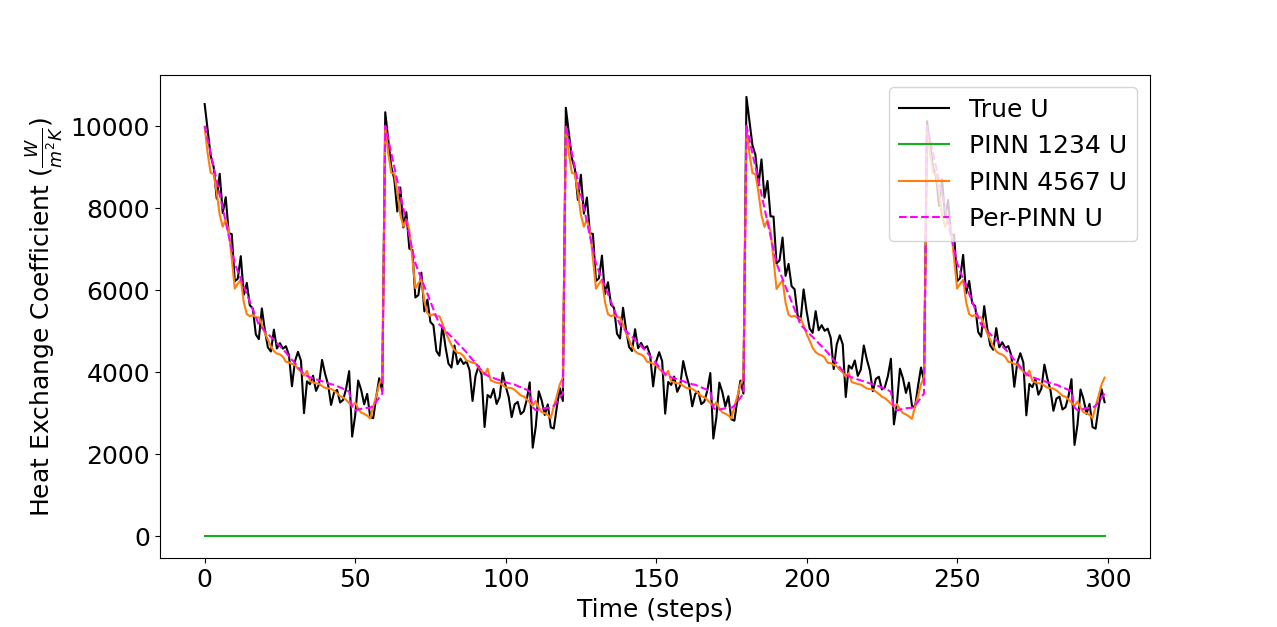}
   \caption{}
   \label{fig:v41} 
\end{subfigure}

\begin{subfigure}[b]{0.45\textwidth}
   \includegraphics[width=1\linewidth]{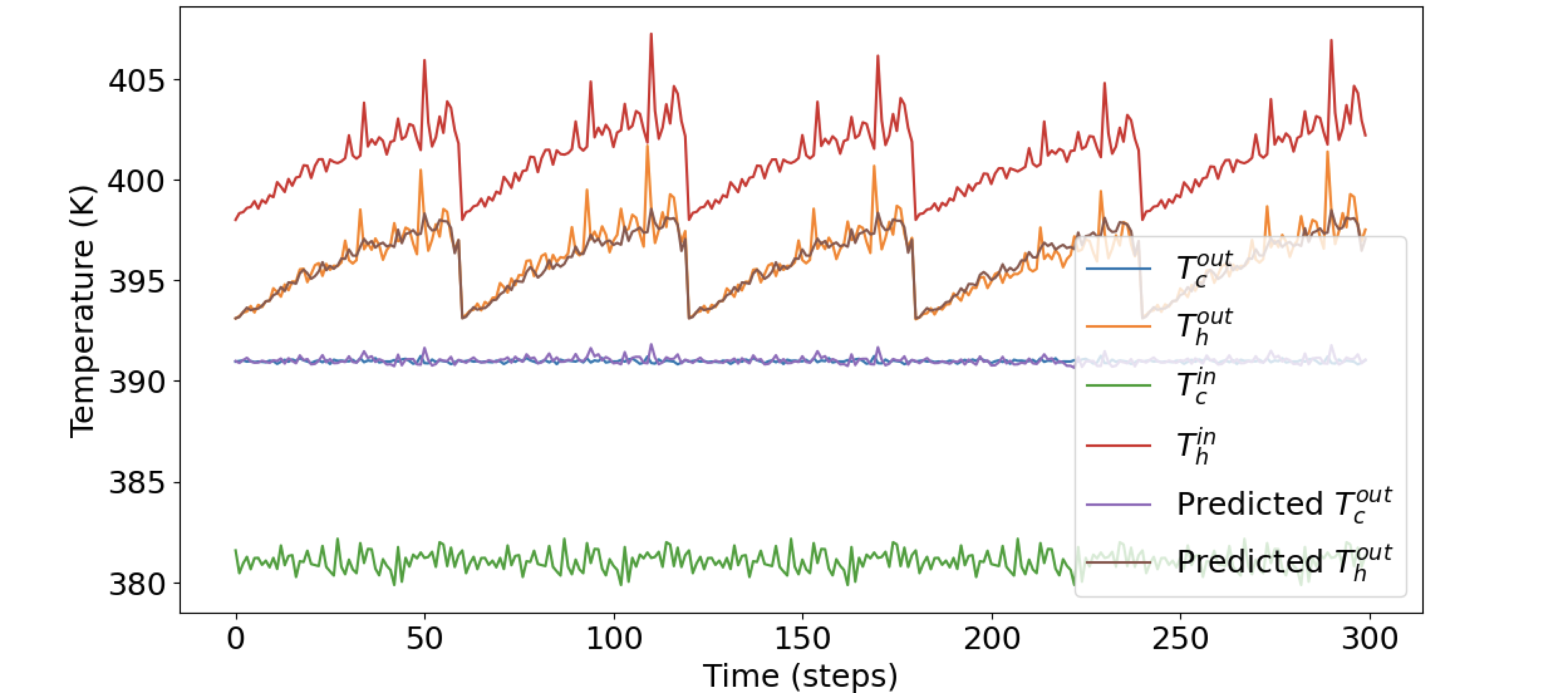}
   \caption{}
   \label{fig:v42}
\end{subfigure}

\caption[Effect of seed on learning $U$]{(a) The learned $U$ for different seeds (b). Prediction of $T_c^{out}$ and $T_h^{out}$ for the PINN $1234$}
\label{fig:Validation4}
\end{figure}

During the simulation studies, it was also observed that the training Per-PINN was much simpler than the traditional PINN. In particular, the training of PINN required a complicated hyperparameter search to provide the best results, and it intrinsically depended on the choice of the seed in the hyperparameter search. We had a choice of $12$ different network structures and two different seeds for each structure (also referred as model), making it $24$ different PINNs searched. Among the $12$ models, we chose the one that had the best training and validation error with closed-loop data (model $9$). Even in this best model, it can be seen that the choice of seed $1234$ versus $4567$ can affect the training and validation errors, as shown in the Table \ref{tab1}. Furthermore, in Figure \ref{fig:Validation4}, it can be seen that for the same network (albeit a different model $5$), different initial seed choices drastically affected the learning of $U(t)$. In particular, in Figure \ref{fig:v41}, it can be seen that the PINN with seed $4567$ was able to learn $U$ properly while the PINN with seed $1234$ was unable to learn $U(t)$ at all. But even without learning $U(t)$, the PINN with seed $1234$ was able to predict $T_c^{out}$ and $T_h^{out}$ correctly, as evidenced in Figure \ref{fig:v42}. 

\begin{figure}
\centering
\begin{subfigure}[b]{0.45\textwidth}
   \includegraphics[width=1\linewidth]{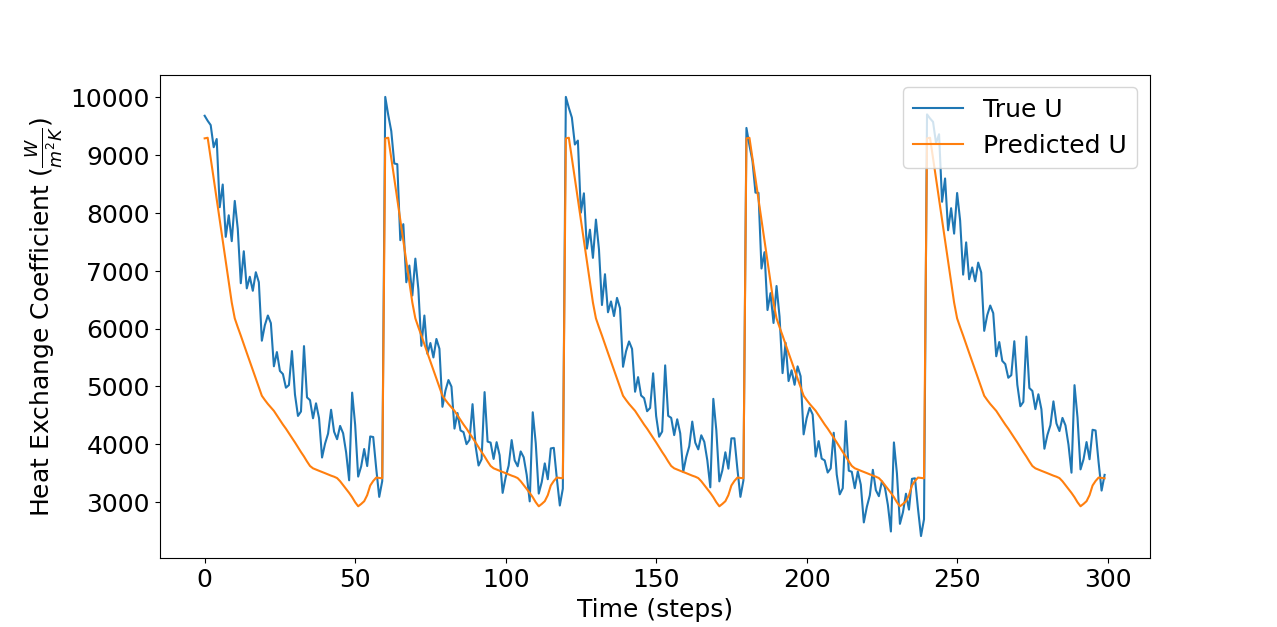}
   \caption{}
   \label{fig:v51} 
\end{subfigure}

\begin{subfigure}[b]{0.45\textwidth}
   \includegraphics[width=1\linewidth]{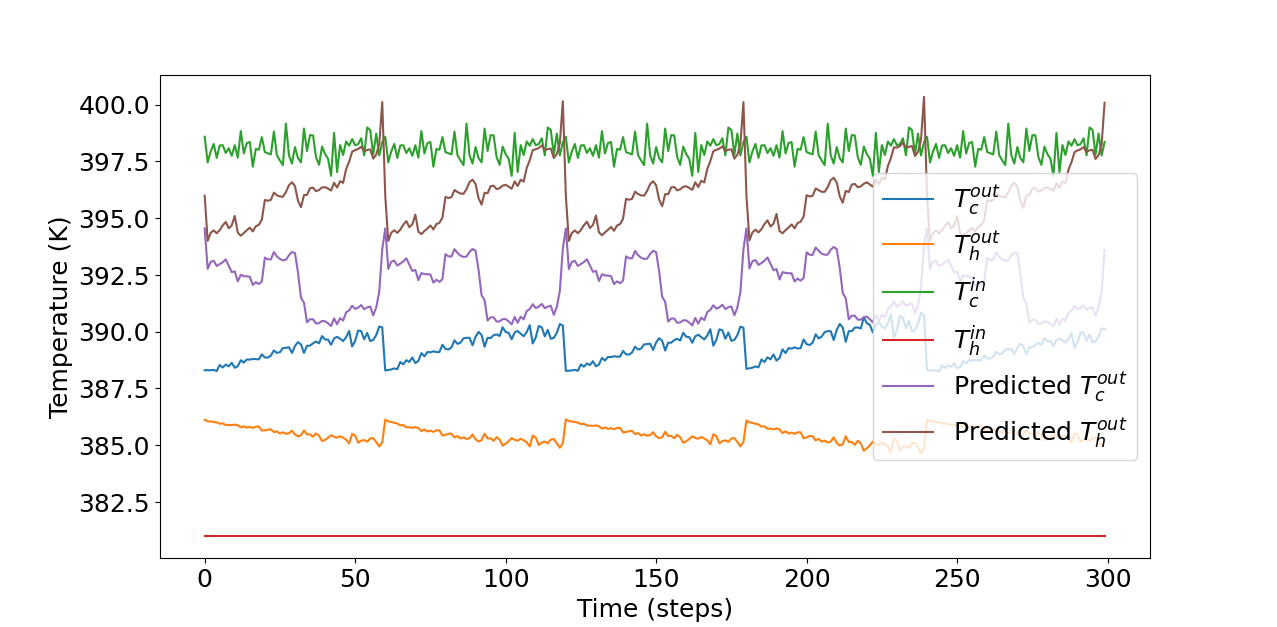}
   \caption{}
   \label{fig:v52}
\end{subfigure}

\caption[PINN producing physically inconsistent data for switched streams]{(a) The learned $U$ (b). Prediction of $T_c^{out}$ and $T_h^{out}$. Note that $T_h^{out}$ (the outlet of the cold stream in switched configuration) has a temperature higher than $T_c^{in}$ (the inlet temperature in switched configuration)}
\label{fig:Validation5}
\end{figure}

Additionally, we see that in one of the trained models (model $6$, which ranked $3^{rd}$ among $12$ in the validation errors) during the hyperparameter search, the resulting PINN violates the energy conservation laws (close to the transition between runs), as illustrated in Figure \ref{fig:Validation5}. This problem is non-existent in Per-PINN,  as it respects physical consistency as a \textit{hard} constraint, unlike PINN which adds the physical laws as a \textit{soft} constraint (such as in an $L_2$ sense).   

\section{Conclusions and Future scope}
In this work, we focused on developing a neural network based framework to model the heat transfer coefficient in a Heat Exchanger with data available from closed-loop operations without any external excitations. The change in heat transfer coefficient due to fouling is considered as an unknown time-varying parameter that drives the system. We first established the identifiability of this time-varying parameter under nominal assumptions on the control structure. Once the identifiability is established, we developed a neural network framework (Per-PINN) which exploits the known ODE model of the HEX to estimate the heat transfer coefficient. This framework is compared to the well-known PINN based framework for parameter estimation where the physical model is used as an additional loss function to the neural network. It is shown through simulations that the Per-PINN performs better than PINN based models, especially when the underlying model structure is known perfectly as the Per-PINN model does not approximate the ODE unlike the PINN. Future investigations would include complexity analysis in training of Per-PINN, applications to time-varying parameter estimation on different classes of ODEs, and extensions to parameter estimation of nonlinear ODEs.

\section*{Appendix A1. Identifiability of $U(t)$}
Given the system \eqref{eq:Id1} identifiability of time varying parameters is established by computing the unknown input reconstructibility matrix and its rank. Starting from the outputs $h_1(x) = T_h^{out} = e_1^Tx, h_2 = T_c^{out} = e_2^Tx$, unknown input degree of reconstructibility is the rank of the matrix 
\[
\mathcal{RM}(h_1,h_2) = \text{rank} \begin{bmatrix}
    \mathcal{L}_{g_1} h_1 \\ \mathcal{L}_{g_1} h_2
\end{bmatrix},
\]
where $\mathcal{L}_{g_1}$ is the Lie derivative operator along the $g_1$ and is given by
\[
\mathcal{L}_{g_1}(h_j) = \frac{\partial h_j}{\partial x}^T g_1 \quad j = 1,2.
\]
The matrix $\mathcal{RM}(h_1,h_2)$ is called the unknown input reconstructibility matrix using functions $h_1,h_2$. Though we used the output functions, in general the unknown input reconstructibility matrix is constructed with the output functions and its finite extensions. We direct the reader to \cite{Martinelli_2022} for detailed explanation about construction of unknown input reconstructibility matrix. Using the dynamics \eqref{eq:Id1}, we compute
\begin{align*}
\mathcal{L}_{g_1}(h_1) &= [1, \ \ 0,\ \ 0]^T \begin{bmatrix}\beta_h (T_c^{out}-T_h^{out}) \\ -\beta_c (T_c^{out}-T_h^{out}) \\ 0   \end{bmatrix} \\
&= \beta_h (T_c^{out}-T_h^{out})\\
\mathcal{L}_{g_1}(h_2) &= [0, \ \ 1,\ \ 0]^T \begin{bmatrix}\beta_h (T_c^{out}-T_h^{out}) \\ -\beta_c (T_c^{out}-T_h^{out}) \\ 0   \end{bmatrix} \\
&= -\beta_c (T_c^{out}-T_h^{out}),
\end{align*}
and 
\begin{align}
\label{eq:UIR}
\mathcal{RM}(h_1,h_2) = \text{rank}\begin{bmatrix}  \beta_h (T_c^{out}-T_h^{out}) \\ -\beta_c (T_c^{out}-T_h^{out})
\end{bmatrix} = 1.
\end{align}
We restate Theorem 1 from \cite{Martinelli_2023} for convenience. 
\begin{theorem}
If the system is observable, then the time-varying parameters of the system are locally identifiable if and only of the unknown input degree of reconstructibility is equal to the number of unknown time varying parameters.    
\end{theorem}
\vspace{.1in}
Since the system \eqref{eq:Id1} observable, and there is only one time varying parameter $U(t)$, and from \eqref{eq:UIR} we know that the unknown input degree of reconstructibility is $1$ we establish the identifiability of $U(t)$. 

\end{document}